\documentclass[a4paper,12pt]{article}
\usepackage{graphicx}
\usepackage{amssymb}
\usepackage{amsmath}
\usepackage{cite}
\usepackage{xcolor}
\def\be{\begin{eqnarray}}
\def\ee{\end{eqnarray}}

\title{Exact solutions for two-body  problems in 1D deformed space with minimal length }
\author{M. I. Samar and V. M. Tkachuk\\ Department for Theoretical Physics, \\ Ivan Franko National University of Lviv,\\ 12 Drahomanov St, Lviv,
UA-79005, Ukraine}
\begin{document}

\maketitle

\begin{abstract}
 We reduce two-body problem  to the one-body problem in general case of deformed Heisenberg algebra leading to minimal length.
 Two-body problems with delta and Coulomb-like interactions are solved exactly. 
 We obtain analytical expression for the energy spectrum for partial cases of deformation function. 
 The dependence of the energy spectrum on the center-of-mass momentum  is found. 
 For special case of deformation function, which correspondes to cutoff procedure in momentum space it is shown that this dependence is more likely to observe for identical particles.



\end{abstract}

\section{Introduction}
Idea of the minimal length have been attracted a lot of attention recently.
Although the first paper on the topic was published in 1947 \cite{Snyder},  interest in
the study increased only in the late 1980s. This interest was motivated by the investigations in  string theory and quantum gravity,
which suggest the existence of the minimal length as a finite lower bound 
to the possible resolution of length \cite{GrossMende,Maggiore,Witten}.
Kempf had shown that such an effect can be achieved by modifying usual canonical
commutation relations \cite{Kempf1994,KempfManganoMann,HinrichsenKempf,Kempf1997}.
The simplest case of the deformed algebra is the following one\cite{Kempf1994}
\be \label{Kempf}
 [\hat{X},\hat{P}]=i\hbar (1+\beta \hat{P}^2),
\ee
leading to minimal length $\hbar\sqrt{\beta}$.

Many one-particle systems were considered in the framework of minimal length hypothesis. 
Exact solutions were found for 
 one-dimensional harmonic oscillator with minimal uncertainty in position \cite{KempfManganoMann} and also 
 with minimal uncertainty in both position and momentum \cite{Tkachuk2003,Tkachuk2004},
  D-dimensional isotropic harmonic oscillator \cite{Chang,Dadic}, 
 three-dimensional Dirac oscillator \cite{Tkachuk2005}, 
 (1+1)-dimensional Dirac oscillator within Lorentz-covariant deformed algebra \cite{Tkachuk2006}, 
a particle in delta potential and double delta potential \cite{Samar1},
 one-dimensional Coulomb-like problem in general case of deformation \cite{Fityo,Samar2} and a particle in the singular inverse square potential \cite{Bouaziz2007,Bouaziz2008}.
 
Also  the perturbation techniques and numerical calculus were applied to different one-particle  quantum systems with minimal length. 
In \cite{Kempf1997} the perturbational D-dimensional isotropic harmonic oscillator was considered. Three-dimensional Coulomb problem with deformed Heisenberg algebra was studied 
within the perturbation theory in the nonrelativistic case \cite{Brau,Benczik,StetskoTkachuk,Stetsko2006,Stetsko2008} 
and in the case of Lorentz-covariant deformed algebra \cite{SamarTkachuk,Samar}. 
The problem of the D-dimensional delta potential in the first order in parameter of deformation is considered in\cite{Ferkous}.
Numerical result for hydrogen atom spectrum in a space with deformed commutation relation was obtained in \cite{Benczik}.

However the description of multiparticle systems is still far from completeness. 
There are only few papers devoted to studies of composite systems ($N$-particle systems) in the deformed
space with minimal length \cite{Buisseret,Tkachuk2010,Tkachuk2012}. In paper \cite{Buisseret} it was shown that ground-state energy of $N$-body can be bounded from below by a  formula
that only requires to know the ground-state energy of a corresponding two-body system.
Authors of   \cite{Tkachuk2010,Tkachuk2012}  proposed an assumptionassumed that  parameter of deformation depends on the mass of particle as
\be \label{g_cond}
\beta=\frac{\gamma}{m^2},
\ee
with $\gamma$ supposing to be some (fundamental) constant for all particles.
This allows to solve a few problems caused by the  minimal length, namely,  problem of violation of the equvalence principle and problem of dependence of kinetic energy on the composition.
Condition (\ref{g_cond}) also explains strangely small result  obtained for the minimal length from a comparison with the observed precession of the perihelion of Mercury\cite{Benczik2002}.
It is interesting that the similar idea,  that particles with different mass feels the deformation in a different way, was successfully applied in case of noncomutative spaces \cite{Gnatenko2013a,Gnatenko2013b,Gnatenko2016,Gnatenko2017}.
This adds to the above idea a more fundamental meaning.

The exact solutions for two-particle eigenproblems have not yet been found. In present paper we consider two-body systems with delta and Coulomb-like interactions in general case of deformed algebra. 
The paper is organized as follows. In section II we brief about general deformed algebra. In section III  we present a reduction of the two-body problem to one-body problem  in general case of deformation.
 In section IV and V, we find the exact solution of two-particle systems with  delta potential and 1D Coulomb-like potential correspondingly. Section VI contains energy spectra for considered systems in some particular cases of deformation. 
 Finally, in section VI we discuss obtained results.

\section{A brief on deformed algebra }
Let us  consider  a  modified  one-dimensional  Heisenberg  algebra which is generated  by
position $\hat{X}$ and momentum $\hat{P}$ hermitian operators satisfying the following relation
\be \label{general_deformation}
 [\hat{X},\hat{P}]=i\hbar f({\hat{P}}),
\ee
where $f$ are called functions of deformation. We assume that it is strictly positive ($f >0$), even function.

We consider a representation leaving position operator undeformed
\be\label{psevdo-position}
&&{\hat{X}}=\hat{x}=i\hbar\frac{d}{dp},\\ \nonumber
&&{\hat{P}=g({p})}.
\ee
From the fact that operators $\hat{X}$ and $\hat{P}$, written in representation (\ref{psevdo-position}), have to satisfy the commutation relation (\ref{general_deformation}), we obtain the following differential equation for  $g(p)$ 
\be\frac{dg(p)}{dp}=f(P), \ee
with $P=g(p).$
Function $g({p})$  is the odd function  defined on $[-b,b]$, with $b=g^{-1}(a)$. Here $a$ represents the limits of momentum $P \in [-a,a]$.
 Minimal length for the deformed algebra is \cite{Maslowski}
\be l_0=\frac{\pi\hbar}{2b}.\label{minimal_length1} \ee
Thus, if $b<\infty$ nonzero minimal length exists and if $b=\infty$ the minimal length is zero. 

We also assume that $b$ is different for different particles. 
According to \cite{Tkachuk2010,Tkachuk2012} we may suppose that $b$
depends on the mass of the particle by the following condition
\be b={\eta}{m}, \ee
with $\eta$  being the same constant for different particles. 

Note that for deformation (\ref{Kempf}) we have $g(p)=\frac{1}{\sqrt{\beta}}\tan(\sqrt{\beta}p)$, $b=\frac{\pi}{2\sqrt{\beta}}$, $l_0=\hbar\sqrt{\beta}$
and  $\eta=\frac{\pi}{2\sqrt{\gamma}}$, with $\gamma$ being the constant introduced in (\ref{g_cond}).

\section{ Two-body problem in deformed space}
Let us assume that in deformed space the two-body Hamiltonian has a similar form as in undeformed one. In the absence of external potential the Hamiltonian reads
\be
\hat{H}= \frac{\hat{P_1}^2}{2m_1}+\frac{\hat{P_2}^2}{2m_2} -V(\hat{X}_1-\hat{X}_2),
\ee
where $V(\hat{X}_1-\hat{X}_2)$
is the interaction potential energy of the two particles.
To express the fact that parameter of deformation is different for different particles we write the representation  for momenta as 
\be &&\hat{P}_i=g(b_i,p_i)=g_i({p_i}), \nonumber \\
&&\hat{X}_i=\hat{x}_i, \label{repr}
\ee with $p_i \in [-b_i,b_i]$ and $ i=1,\,2 $ enumerates the particles. It is also natural to suppose that the operators corresponding to different particles commute with one another. Without loss of generality we assume $b_1\geq b_2$, where equality is achieved in case of identical particles.

Using representation (\ref{repr}) the Hamiltonian can be written as the following
\be
\hat{H}=\frac{1}{2M} \left[\frac{g_1^2(p_1)}{\mu_1}+\frac{g_2^2(p_2)}{\mu_2}\right] -V(\hat{x}_1-\hat{x}_2).
\ee
Here we use notation  $\mu_1=\frac{m_1}{M}$,  $\mu_2=\frac{m_2}{M}$ and 
$M=m_1+m_2$.

Let us introduce the  center-of-mass coordinate and momentum  $\hat{x}_0$ and ${p}_0$ and relative-motion ones $\hat x$ and $p$ in the traditional way
\be
&&\hat x_0=\mu_1 \hat x_1+\mu_2 \hat x_2,  \ \ \ \ \ \ \ \ \ \ \ \       p_0=p_1+p_2,\\
&&\hat x=\hat x_1- \hat x_2,  \ \ \ \ \ \ \ \ \ \ \ \ \ \ \ \ \ \ \ \    p=\mu_2p_1-\mu_1 p_2.
\ee

The inverse transformation reads
\be
&&\hat x_1=\hat x_0+\mu_2 \hat x,  \ \ \ \ \ \ \ \ \ \ \ \  \ \ \      p_1=\mu_1p_0+p,\\
&&\hat x_2=\hat x_0-\mu_1\hat x,  \ \ \ \ \ \ \ \ \ \ \ \  \ \ \      p_2=\mu_2p_0-p.
\ee
Note that $p_0$ is an integral of motion, because it commutes with the Hamiltonian.

\begin{figure}[h!]
\centering
\includegraphics[width=10cm]{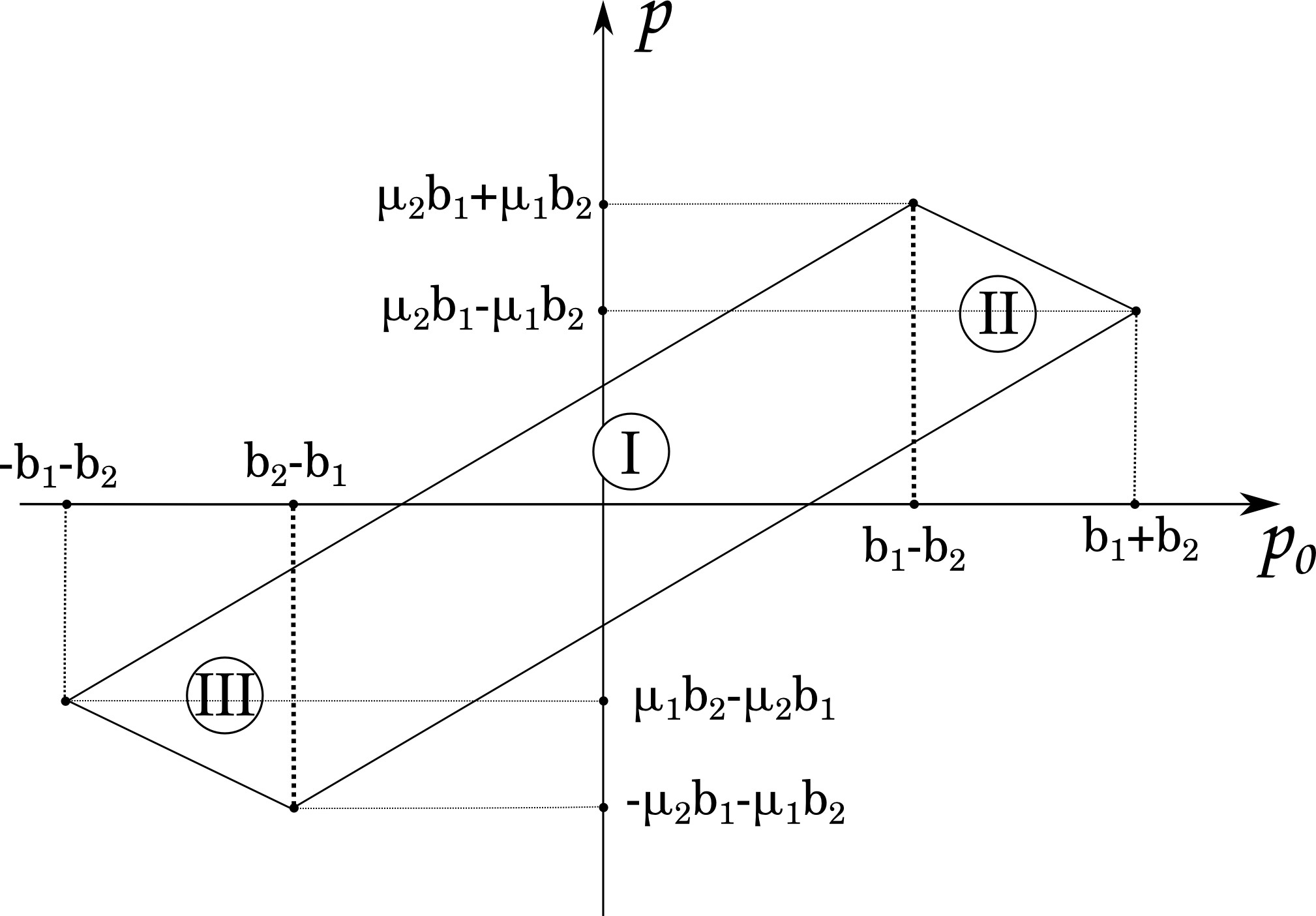}
\caption{\footnotesize{Two-particle momentum space}}
\label{fig.1}
\end{figure}

 The two-particle momentum space after change of variables is presented in Fig. \ref{fig.1}.
As it can be seen in Fig. \ref{fig.1}  range of relative-motion momentum  is differrent depending on the value of the momentum of the center-of-mass $p_0$. 

Results of the analysis of possible ranges for center-of mass and relative-motion momenta  are presented in Table \ref{tab}. We see that there are three different domains in the momentum space, which are set by the value of $p_0$. Relative-motion momentum belongs to the asymmetric region $[c_1,c_2]$, were values of $c_1$ and $c_2$ depend on the value of the momentum of the center-of-mass $p_0$.
\begin{table}[h!]
\caption{Ranges of momenta for different domains\label{tab}}
\centering
\begin{tabular}{|c|c|c|c|c|}
\hline
Domain &range of $ p_0$ &range of $p \in [c_1,c_2]$ \\
\hline
I& $[-(b_1-b_2),b_1-b_2]$&$[-(b_2-\mu_2p_0) ,b_2+\mu_2 p_0]$\\
\hline
II& $[(b_1-b_2), b_1+b_2]$&$[-(b_2-\mu_2p_0), b_1-\mu_1 p_0]$\\
\hline
III& $[-(b_1+b_2),-(b_1-b_2)]$&$[-(b_1+\mu_1p_0), b_2-\mu_2 p_0]$\\
\hline
\end{tabular}
\end{table}

Rewriting the Hamiltonian in terms of center-of-mass and relative-motion variables we obtain
\be
\hat{H}=\frac{1}{2M} \left[\frac{g_1^2(\mu_1p_0+p)}{\mu_1}+\frac{g_2^2(\mu_2p_0-p)}{\mu_2}\right] +V(\hat{x}).
\ee
Schr\"odinger equation in momentum representation for two-particle system can be written in a similar way as it was done for one-particle systems \cite{Samar1, Samar2}:
\be\label{main}
\frac{1}{2M}G^2(p_0,p)\phi(p)+\int_{c_1}^{c_2}U(p-p')\phi(p')dp'=E\phi(p),
\ee
where $U(p-p')$ is the  kernel of potential energy operator and 
\be\label{g}
G(p_0,p)^2=\frac{g_1^2(\mu_1p_0+p)}{\mu_1}+\frac{g_2^2(\mu_2p_0-p)}{\mu_2}.
\ee
The main  difference of equation (\ref{main}) from  the corresponding one for one-particle systems  is that  $G(p_0,p)$ depends on $p_0$ and  that the endpoints of $p$, denoted as $c_1$ and $c_2$,  now is not symetric in general and also depends on $p_0$. Note that $p_0$ is a constant of motion.

Thus, two-body problem in deformed space can be reduced to one-body problem.

\section{Delta function interaction in deformed space with minimal length}
In undeformed case attractive  delta potential in the coordinate representation 
\be
V(x)=-2\pi\hbar U_0\delta(x)
\ee
corresponds to constant potential in momentum space 
\be \
U(p-p')=-U_0.
\ee
We assume that in deformed space delta potential is still expressed by constant potential in momentum space. Schr\"odinger equation (\ref{main}) then reads
\be \label{Schrodinger_Delta}
\frac{1}{2M}G^2(p_0,p)\phi(p)-U_0\int_{c_1}^{c_2}\phi(p')dp'=E\phi(p).
\ee
The solution of (\ref{Schrodinger_Delta}) can be easily obtained
\be\label{phi}
\phi(p)=\frac{2M U_0\tilde{\varphi}}{G^2(p_0,p)+ q^2}.
\ee
Here we use the notation
\be
\tilde{\varphi}=\int_{c_1}^{c_2}\phi(p')dp',
\ee
and
\be
q=\sqrt{-2ME}.
\ee
Integrating (\ref{phi}) over $p$ in the range $[c_1, c_2]$, we obtain
\be\label{delta_energy_cond}
2M U_0\int_{c_1}^{c_2}\frac{dp}{G^2(p_0,p)+q^2}=1,
\ee
with $g(p)$ presented in (\ref{g}).
Note that the limits of integration $c_1$ and $c_2$ depends on the value of the center-of-mass momentum $p_0$ and are presented in Table \ref{tab}.

\section{ Coulomb-like interaction in deformed space with minimal length }
The kernel of the potential energy operator in momentum representation  in undeformed case has the following form \cite{Samar2}
\be\label{kernelnondef}
U(p-p')=-\frac{\alpha}{2\hbar}(2 i \theta(p'-p)-i+\cot{(\pi\delta)}),
\ee
with $\theta(p'-p)$ being the Heaviside step function $\delta \in [0,1) $. Note that the endpoint $1$ is excluded because in the limit of $\delta$ to $0$, we obtain the same definition of potential energy operator as in the limit of $\delta$ to $1$.

We assume that the kernel of potential energy operator remains unchanged in deformed space, which means that inverse position operator can be written as the following
\be\label{1/X}
\frac{1}{\hat{X}}\phi(p)=-\frac{i}{\hbar}\int_{c_1}^{p}\phi(p')dp'+\frac{i+\cot{(\pi\delta)}}{2\hbar}\int_{-c_1}^{c_2}\phi(p')dp'.\ee
 The inverse position operator defined above is linear and in the limit of  $c_1$ to $-\infty$  and $c_2$ to $\infty$ coincides with the undeformed one.
The Schr\"odinger equation for 1D Coulomb-like problem in deformed space with minimal length reads
\be\label{Shroed_1/X}
\frac{1}{2M}G(p_0,p)^2\phi(p)-\frac{\alpha}{2\hbar}\left[(i+\cot(\pi\delta))\int_{c_1}^{c_2}\phi(p')dp'-2i\int_{c_1}^{p}\phi(p')dp'\right]=E\phi(p). \label{eigenequation}
\ee
Differentiating the latter integral equation, we obtain the differential one
\be\label{f44}
\frac{1}{2M} \left(G^2(p_0,p)\phi(p)\right)'+\frac{i\alpha}{\hbar}\phi(p)=E\phi'(p),
\ee
 which can be written as 
\be\label{f45}
\left[ \left(G^2(p_0,p)+q^2\right)\phi(p)\right]'=\frac{2iM\alpha}{\hbar}\phi(p), 
\ee
with  $q=\sqrt{-2ME} $.
The solution of (\ref{f45}) reads
\be
\phi(p)=\frac{C}{G^2(p_0,p)+q^2}e^{-{i}\varphi(p)}.\label{eigenfunction}
\ee
Here normalization constant is
\be C=\left(\int_{-c_1}^{c_2}\frac{dp'}{(G^2(p_0,p')+q^2)^2}\right)^{-\frac{1}{2}}.\ee
We also use notation
\be
\varphi(p)=\frac{2M\alpha}{\hbar}\int_0^p\frac{dp'}{G^2(p_0,p')+q^2}.
\ee
The eigenfunction (\ref{eigenfunction}) is the solution of equation (\ref{f45}) but not necessarily (\ref{Shroed_1/X}). Only the eigenfunctions  (\ref{eigenfunction}) with some specific values of energy $E$ would satisfy equation (\ref{Shroed_1/X}). To find these energies let us calculate the integrals from  (\ref{eigenequation}). Using formulas (\ref{f45}) and (\ref{eigenfunction}), we have
\be\label{I1}
\int_{c_1}^{c_2}\phi(p')dp'=\frac{i\hbar C}{2M\alpha}\left(e^{-i\varphi(c_2)}-e^{-i\varphi(c_1)}\right),
\ee
\be\label{I2}
\int_{c_1}^{p}\phi(p')dp'=\frac{i\hbar C}{2M\alpha}\left(e^{-i\varphi(p)}-e^{-i\varphi(c_1)}\right).
\ee
Substituting obtained results (\ref{eigenfunction}), (\ref{I1}) and (\ref{I2}) into equation (\ref{eigenequation}), we find 
\be \label{sin1}
\sin\left(\frac{\varphi(c_2)-\varphi(c_1)}{2}-\delta\pi\right)=0.
\ee 
Finally, energy spectrum can be found from
\be\label{energy_condition}
\frac{M\alpha}{\hbar}\int_{c_1}^{c_2}\frac{dp}{G^2(p_0,p)+q^2}=\pi(n+\delta), 
\ee
with $n=0,1,\ldots$ and $\delta \in [0,1)$. The limits of integration $c_1$ and $c_2$ depend on the value of the center-of-mass momentum $p_0$ and are presented in Table \ref{tab}. 

Hence, the exact solution for two-particle problem with Coulomb-like interaction is found in general case of deformed Heisenberg algebra with minimal length. 

\section{Energy spectra for particular cases of deformation}

Condition for energy spectrum for Coulomb-like problem  (\ref{energy_condition}) contains the same integral as in the corresponding condition for delta potential (\ref{delta_energy_cond}). Therefore finding the spectrum for particular cases of deformation function is similar for the both systems.
We combine conditions (\ref{energy_condition}) and (\ref{delta_energy_cond}) into the following one 
\be
\label{Cond}
\int_{-c_1}^{c_2}\frac{dp}{G^2(p_0,p)+q^2}=\frac{\kappa}{M},
\ee
with $\kappa=\frac{1}{2U_0}$ for delta potential and $\kappa=\frac{\pi\hbar(n+\delta)}{\alpha}$ for Coulomb-like one.

\textbf{Example 1.} 

Let us consider the simplest deformed commutation relation leading to minimal length and corresponding to ultraviolet cutoff 
\be \label{cutoff}
&&f_i(P_i)=1,\, P_i\in[-b_i,b_i],\\
&&g_i(p_i)=p_i, \,  p_i\in\left[-b_i,b_i\right],
\ee
with  $i=1,2$ enumerates the particles.
In this case energy condition on energy spectrum  (\ref{Cond}) is
\be
\label{delta_cond}
\int_{-c_1}^{c_2}\frac{dp}{\frac{(\mu_1p_0+p)^2}{\mu_1}+\frac{(\mu_2p_0-p)^2}{\mu_2}+q^2}=\frac{\kappa}{M}.
\ee
Taking the integral in the latter equation we obtain
\be\label{1}
 \sqrt{\frac{\mu_1\mu_2}{p_0^2+q^2}}\left(\arctan{\left(\frac{c_2}{p_0^2+q^2}\right)}-\arctan{\left(\frac{c_1}{p_0^2+q^2}\right)}\right)=\frac{\kappa}{M}.
\ee
In the case of $|p_0|\ll b_1-b_2$ (domain I) from (\ref{1}) we obtain the energy spectrum in a series over small $\frac{1}{b_2}$ 
\be
E=\frac{p_0^2}{2M}-\frac{\pi^2\mu}{2\kappa^2}+\frac{2\pi^2\mu^2}{\kappa^3b_2}-\frac{6\pi^2\mu^3}{\kappa^4b_2^2}+O\left(\frac{1}{b_2^3}\right),
\ee
with $\mu=\mu_1\mu_2M$ denoting the reduced mass.
Dependence  of the energy spectrum on $p_0$ is hidden in the higher order correction over $\frac{1}{b_2}$.
It is  interesting that in this case correction to the energy spectrum depends only on the parameter of deformation $b_2$  of the particle, which feels the deformation stronger than the other one.

The energy spectrum of two-particle system with delta function interaction similarly to undeformed space consists of one energy level
\be
E=E^0+\frac{16\pi^2\mu^2U_0^3}{b_2}-\frac{96\pi^2\mu^3U_0^4}{b_2^2}+O\left(\frac{1}{b_2^3}\right).
\ee
The spectrum for the system with Coulomb-like interaction is
\be
E_n=E_n^0+{\frac {2{\mu}^{2}{\alpha}^{3}}{\pi \,{
h}^{3} \left(n+\delta\right)^3 b_2 } }-{\frac {6\mu^3{\alpha}^{4}}{{\hbar}^{4}{
\pi }^{2} \left( n+\delta \right)^4 b_2^{2}}}
+O\left(\frac{1}{b_2^3}\right).
\ee
Here \be E^0=\frac{p_0^2}{2M}-{2\pi^2\mu U_0^2}\ee  and \be E_n^0=\frac{p_0^2}{2M}-{\frac {{\alpha}^{2}\mu}{2{h}^{2} \left(n+\delta \right)^{2} }}\ee denote undeformed energy spectra of two-particle system with delta and Coulomb-like interaction, respectively.

Some different results we obtain for identical particles ($m_1=m_2=m$ and $b_1=b_2=b$).
 In such a situation domain I on fig. \ref{fig.1} vanishes and domains II and III can be combined  in single one with $p \in [-2b+|p_0|,2b-|p_0|]$. 
Equation (\ref{delta_cond}) yields
\be
 \frac{1}{\sqrt{p_0^2+q^2}}\arctan{\left(\frac{2b-|p_0|}{\sqrt{p_0^2+q^2}}\right)}=\frac{\kappa}{M}.
\ee
Finally, energy spectrum expansion up to order $1/b^2$ writes
\be \label{energy}
E=\frac{p_0^2}{2M}-\frac{\pi^2\mu}{2\kappa^2}+\frac{2\pi^2\mu^2}{\kappa^3b}-\frac{6\pi^2\mu^3}{\kappa^4b^2}\left(1-\frac{2}{3}\frac{\kappa|p_0|}{M}\right).
\ee

Equation (\ref{energy}) for delta and Coulomb-like interactions  is 
\be
E=E^0+\frac{16\pi^2\mu^2U_0^3}{b}-\frac{96\pi^2\mu^3U_0^4}{b^2}\left(1-\frac{|p_0|}{3MU_0}\right)
\ee
and
\be
E_n=E^0_n+{\frac {2{\mu}^{2}{\alpha}^{3}}{\pi \,{
h}^{3} \left(n+\delta\right)^3 b } }-{\frac {6\mu^3{\alpha}^{4}}{{\hbar}^{4}{
\pi }^{2} \left( n+\delta \right)^4 {{ b}}^{2}}}\left(1-\frac{2\pi\hbar(n+\delta)|p_0|}{3M\alpha}\right)
\ee
correspondingly.

From (\ref{energy}) we conclude that in the case of identical particles dependence on $p_0$ is present in second order correction over $\frac{1}{b}$, while for different particles this dependence is present in higher order corrections. Therefore, effect of dependence of the energy spectrum on $p_0$  of the considerable problem is more likely to observe for identical particles.

\textbf{Example 2}

Another algebra for which analytical results were obtained for identical particles $(\mu_1=\mu_2=1/2)$  is given by deformation function
\be
&&f_i(P_i)=(1+\beta_i P_i^2)^{3/2},  \ \  P\in(-\infty,\infty), \\
&&g_i(p_i)=\frac{p_i}{\sqrt{1-\beta_i p_i^2}}, \ \  p_i\in\left[-b_i,b_i\right], b_i=\frac{1}{\sqrt{\beta_i}}
\ee
The identical particles feels the deformation in the same way, that is why we 
consider $\beta_1=\beta_2=\beta$.
Condition on the energy spectrum can be written as
\be
\label{delta_cond2}
\int_{-c_1}^{c_2}\frac{dp}{\frac{(\mu_1p_0+p)^2}{(1-\beta (\mu_1 p_0+p)^2 )\mu_1}+\frac{(\mu_2p_0-p)^2}{(1-\beta (\mu_2 p_0+p)^2 )\mu_2}+q^2}=\frac{\kappa}{M}.
\ee
The series of energy spectrum  over   $\sqrt{\beta}$  up to the second order writes
\be \label{energy_equal}
E=\frac{p_0^2}{2M}-{\frac {{\pi }^{2}\mu}{{2\kappa}^{2}}}+{\frac {{4\pi }^{2}{\mu}^{2
}\sqrt {\beta}}{{\kappa}^{3}}}-\left({\frac {{{p_0}}^{4}}{32\mu}}+{\frac {3\mu{
\pi }^{2}{{p_0}}^{2}}{{4\kappa}^{2}}}-{\frac {3{\mu}^{3}{\pi }^{4}}{{
2\kappa}^{4}}}-{\frac {24{\mu}^{3}{
\pi }^{2}}{{\kappa}^{4}}}
\right)\beta.\nonumber \\
\ee
Thus, the energy spectrum for two-particle system with delta  and Coulomb-like interaction writes
\be \label{energy_equal}
E=E^0+{32\pi^2\mu^2U_0^3}\sqrt{\beta}-\left(\frac{p_0^4}{32\mu}+3\pi^2\mu U_0^2p_0^2-24\pi^4\mu^3U_0^4-384\pi^2\mu^3U_0^4\right)\beta \nonumber \\
\ee
and
\be \label{energy_equal}
E_n=E^0_n+{\frac{4{\mu}^{2
}\alpha^3\sqrt {\beta}}{\pi\hbar^{3}(n+\delta)^3}}-\left({\frac {{{p_0}}^{4}}{32\mu}}+{\frac {3\mu{
\alpha }^{2}{{p_0}}^{2}}{4\hbar^{2}(n+\delta)^2}}-\frac {3{\mu}^{3}{\alpha }^{4}}{
2\hbar^{4}(n+\delta)^4}-{\frac {24{\mu}^{3}{
\alpha }^{4}}{\pi^2\hbar^{4}(n+\delta)^4}}
\right)\beta \nonumber \\
\ee
respectively.
Similarly to the previous example dependence on the center-of-mass momentum is not presented in the leading correction. However in the next order correction to the energy we have dependence on center-of-mass momentum as $p_0^2$ and $p_0^4$ unlike the first example of deformation, where we obtain the second order correction proporional to $|p_0|$.

\section{Conclusion}
In this paper we have studied two-particle problem in the general case of deformed Heisenberg  algebra leading to the minimal length.
We have assumed that different particles feel the deformation in different ways.
By introducing the center-of-mass and relative-motion coordinates and momenta we succeeded to reduce  two-particle problem to the one-particle one. 
The pecularity of the obtained one-particle problem is in it dependence on the center-of-mass momentum.
 
Two-particle system with delta interaction as well as Coulomb-like one have been considered.
We have obtained exactly the wave functions and energy spectra of the mentioned problems in general case of deformed algebra. 
We have cosidered two  partial examples of deformation function. 
Expanding the energy spectrum over parameter of deformation we conclude that the dependence on the center-of-mass momentum is not present in the main correction to the energy for both considered examples.
For the deformation function that correspondes to ultaviolet cutoff we have obtained that for different particles dependence on the center-of-mass momentum is also not present in the second order correction, although for identical particles second order correction is proportional to $|p_0|$.
This means that for considered type of deformation the effect of dependence of energy spectrum on the center-of-mass momentum  is more likely to observe for  identical particles. In the other example of deformation we have obtained for identical particles that the second order correction to the energy has terms  proportional to  $p_0^2$ and $p_0^4$. It means that in different cases of deformation dependence on  center-of-mass momentum is essentially  different.

\section{Acknowledgement}
This work was supported in part by the European Commission under the project STREVCOMS PIRSES-2013-612669 and project FF-30F (No. 0116U001539) from the
Ministry of Education and Science of Ukraine.
We also would like to thank to Dr. Khrystyna Gnatenko for careful
reading of the manuscript.


\begin{thebibliography}{999}
\bibitem{Snyder}  H. S. Snyder, Phys. Rev. 71, 38 (1947).
\bibitem{GrossMende}D. J. Gross  and P. F. Mende,  Nucl. Phys. B 303, 407 (1988).
\bibitem{Maggiore}M. Maggiore,  Phys. Lett. B 304, 65 (1993).
\bibitem{Witten}E. Witten,  Phys. Today 49, 24 (1996).
\bibitem{Kempf1994}A. Kempf, J. Math. Phys. 35, 4483 ( 1994).
\bibitem{KempfManganoMann} A. Kempf, G. Mangano  and R. B. Mann, Phys. Rev. D 52, 1108 (1995).
\bibitem{HinrichsenKempf}H. Hinrichsen  and A. Kempf, J. Math. Phys. 37, 2121 (1996).
\bibitem{Kempf1997} A. Kempf, J. Phys. A 30, 2093 (1997).


\bibitem{Tkachuk2003}C. Quesne and V. M. Tkachuk, J. Phys. A 36, 10373 (2003).
\bibitem{Tkachuk2004}C. Quesne and V. M. Tkachuk, J. Phys. A 37, 10095 (2004).
\bibitem{Chang} L. N. Chang, D. Minic, N. Okamura and T. Takeuchi, Phys. Rev. D 65, 125027 (2002).
\bibitem{Dadic} I. Dadic, L. Jonke and S. Meljanac, Phys. Rev. D 67, 087701 (2003).
\bibitem{Tkachuk2005} C. Quesne and V. M. Tkachuk, J. Phys. A 38, 1747 (2005).
\bibitem{Tkachuk2006} C. Quesne and V. M. Tkachuk, J. Phys. A 39, 10909 (2006).
\bibitem{Samar1} M. I. Samar, V. M Tkachuk, J. Math. Phys. 57, 042102 (2016).
\bibitem{Samar2} M. I. Samar, V. M Tkachuk, J. Math. Phys. 57, 082108 (2016).
\bibitem{Fityo} T. V. Fityo, I. O. Vakarchuk and V. M. Tkachuk, J. Phys. A 39, 2143 (2006).
\bibitem{Bouaziz2007}D. Bouaziz and M. Bawin, Phys. Rev. A 76, 032112 (2007).
\bibitem{Bouaziz2008} D. Bouaziz and M. Bawin, Phys. Rev. A 78, 032110 (2008).

\bibitem{Brau} F. Brau, J. Phys. A 32, 7691 (1999).
\bibitem{Benczik}S. Benczik, L. N. Chang, D. Minic and T. Takeuchi, Phys. Rev. A 72, 012104 (2005).
\bibitem{StetskoTkachuk}M. M. Stetsko and V. M. Tkachuk, Phys. Rev. A 74, 012101 (2006).
\bibitem{Stetsko2006} M. M. Stetsko, Phys. Rev. A 74, 062105 (2006).
\bibitem{Stetsko2008}M. M. Stetsko and V. M. Tkachuk, Phys. Lett. A 372, 5126 (2008).
\bibitem{SamarTkachuk}M. I. Samar and V. M Tkachuk, J. Phys. Stud. 14, 1001 (2010).
\bibitem{Samar} M. I. Samar, J. Phys. Stud. 15, 1007 (2011).
\bibitem{Ferkous} N. Ferkous, Phys. Rev. A 88, 064101 (2013).

\bibitem{Buisseret}  F. Buisseret, Phys. Rev. 82, 062102 (2010).
\bibitem{Tkachuk2010} C. Quesne and V. M. Tkachuk, Phys. Rev. A 81,012106 (2010).
\bibitem{Tkachuk2012} V. M. Tkachuk, Phys. Rev. A 86, 062112 (2012).

\bibitem{Benczik2002} S. Benczik, L. N. Chang et al., Phys. Rev. D 66, 026003 (2002).

\bibitem {Gnatenko2013a} Kh. P. Gnatenko, Phys. Lett. A 377,  3061 (2013). 
\bibitem{Gnatenko2013b} Kh. P. Gnatenko,  J. Phys. Stud. 17,  4001  (2013). 
\bibitem{Gnatenko2016}Kh. P. Gnatenko, V. M. Tkachuk, Mod. Phys. Lett. A 31, No. 5, 1650026 (2016). 
\bibitem{Gnatenko2017}Kh. P. Gnatenko, V. M. Tkachuk, Phys. Lett. A 381, 2463 (2017). 
\bibitem{Maslowski} T. Maslowski, A. Nowicki and V. M. Tkachuk, J. Phys. A 45, 075309 (2012).
\end{thebibliography}
\end{document}